\documentclass[12pt,a4paper]{article}

\usepackage{graphicx}
\usepackage{mathletters}
\usepackage[footnotesize,bf]{caption} 

\newcommand{\lesssim}{\hbox{\rlap{$^<$}$_\sim$}}

\begin{document}

\title{
  Spherical Collapse of a Mass-Less Scalar Field
  With Semi-Classical Corrections}


\author{ 
  Shai Ayal\thanks{email: shai@shemesh.fiz.huji.ac.il}
  \ \&
  Tsvi Piran\thanks{email: tsvi@shemesh.fiz.huji.ac.il} \\
  \small 
  \textit{
    Racah Institute for Physics, The Hebrew University, 
    Jerusalem, Israel, 91904
    }
  } 
\date{}
\maketitle

\begin{abstract}
  We investigate numerically spherically symmetric collapse of a
  scalar field in the semi-classical approximation. We first verify
  that our code reproduces the critical phenomena (the Choptuik
  effect) in the classical limit and black hole evaporation in the
  semi classical limit. We then investigate the effect of evaporation
  on the critical behavior. The introduction of the Planck length by
  the quantum theory suggests that the classical critical phenomena,
  which is based on a self similar structure, will disappear. Our
  results show that when quantum effects are not strong enough,
  critical behavior is observed. In the intermediate regime,
  evaporation is equivalent to a decrease of the initial amplitude. It
  doesn't change the echoing structure of near critical solutions. In
  the regime where black hole masses are low and the quantum effects
  are large, the semi classical approximation breaks down and has no
  physical meaning.
\end{abstract}

\newpage

\section{Introduction}
Spherical symmetric collapse of a scalar field was studied
analytically by Christodoulou \cite{christ} who concluded that for
near trivial initial data, the field disperses to future null infinity
leaving an empty space-time, and for ``stronger'' initial data, the
field implosion forms a black-hole. This result was verified
numerically by Goldwirth \& Piran \cite{goldwirth}.  If we now
characterize the ``strength'' of the initial data by some well defined
parameter $p$ which is monotonously growing with the ``strength''
(e.g. the initial amplitude of the field), we can expect that there
will exist a critical value $p^*$ so that data with $p<p^*$ disperses
while data with $p>p^*$ creates a black-hole. An intriguing question
is what happens at $p=p^*$. This question was investigated numerically
by Choptuik \cite{chop}. His findings were
\begin{itemize}
\item In the limit $p \rightarrow p^*$ the asymptotic behavior of the
  scalar field and the metric is {\em universal} - independent of the
  initial profile of the field. The field and metric also posses a
  symmetry - discrete self similarity (DSS) in the neighborhood of the
  critical solution's horizon.
\item The DSS is a periodical behavior at the origin with
  exponentially smaller periods in proper time: $\tau \mapsto \tau
  e^{-\Delta}$ where $\tau=t^*-t$, $t^*$ being the time of formation
  of the critical black-hole, $\Delta\approx 3.44$.
\item The mass of the black hole formed by marginally {\em
    super}critical data has a scaling law $M \propto (p-p^*)^\gamma$
  with $\gamma \approx 0.375$.
\end{itemize}

Perhaps one of the striking conclusions from these findings is that it
is possible to produce a black-hole with an infinitesimal mass. The
limiting case zero mass black-hole, the ``choptuon'' can be viewed as
a collapsing ball of field energy for which the rate of collapse is
exactly balanced by the energy loss by radiation, so that when the
ball shrinks to zero radius all of its energy is radiated away
\cite{nrg}. The choptuon is actually a naked singularity, visible from
null infinity \cite{stewart} although it is not generic and it is
destroyed by an arbitrarily small perturbation.

Hawking \cite{hawking} has shown that within the semi-classical
approximation black holes evaporate thermally.  The rate of
evaporation depends on the black hole's mass via the relation
\cite{page}:
\begin{equation}
  \frac{\partial M}{\partial t} \propto -M^{-2}, \label{eq:evapor}
\end{equation}
with the proportionality constant depending on the number of particle
species that can be emitted by the black-hole at a given temperature.
Integration of this relation yields a black-hole lifetime which is
proportional to $M^3$. This suggests that quantum effects will change
the Choptuik effect since infinitesimally small black holes evaporate
almost instantly.

The semi-classical approach to gravitation stipulates that we can
write the semi-classical Einstein equation \cite{semiclass}
\begin{equation}
        G_{\mu\nu}=8\pi{\langle T^Q\rangle}_{\mu\nu} ,
\end{equation}
where $\langle T^Q\rangle$ is the expectation value of the effective
quantum energy-momentum tensor.  This approach has been tested to give
rise to evaporation of the black hole satisfying Eq.\ 
(\ref{eq:evapor}) in \cite{piran}.

We explore the evaporation of a black hole formed by collapse of a
scalar field and the effect of the evaporation on the Choptuik
phenomena. To this end, we investigate, numerically, the spherical
collapse of a scalar field with the addition of an appropriate
$\langle T^Q\rangle$. We use $\langle T^Q\rangle$ adapted from a 2D
expectation value that reproduces the Hawking radiation \cite{piran}.
This in turn forces us to solve the evolution equations.  This is
significantly harder in our coordinates then the usual method of
solving the constraint equations used in \cite{garf} and
\cite{stewart}. In fact, this is the first time that a numerical
solution using the evolution equations has been presented.  Another
problem with the above $\langle T^Q\rangle$ is that the tensor is
divergent at the origin. We renormalize the tensor to bypass this
problem. This renormalization violates the Bianchi identity near the
origin however the semi classical approximation is not valid there
anyway. The code is tested with the classical problem to reproduce the
Choptuik effect. After this we check black hole evaporation.  Finally
we examine the effect of evaporation on the Choptuik phenomena.

The introduction of quantum effects via $\langle T^Q\rangle$
immediately introduces the Planck length scale denoted here by
$\sqrt{\alpha}$. The addition of a length scale to the previously
scale-less problem of the massless scalar field suggests that the
critical phenomena present there will disappear \cite{chiba}. However,
our results show that the critical phenomena does not disappear within
the regime of validity of the semi-classical approximation.

We discuss the problem of classical \& semi-classical scalar field
collapse in Sec. II \& III. Our numerical scheme is described in Sec.
IV and our results in Sec. V.  In Sec. VI we summarize our findings.
Finally in Appendix A we derive several auxiliary quantities needed
for the numerical scheme.

\section{Classical Scalar Field Collapse}
\begin{figure}
  \centering
  \noindent
  a 
  \includegraphics[width=3.9cm]{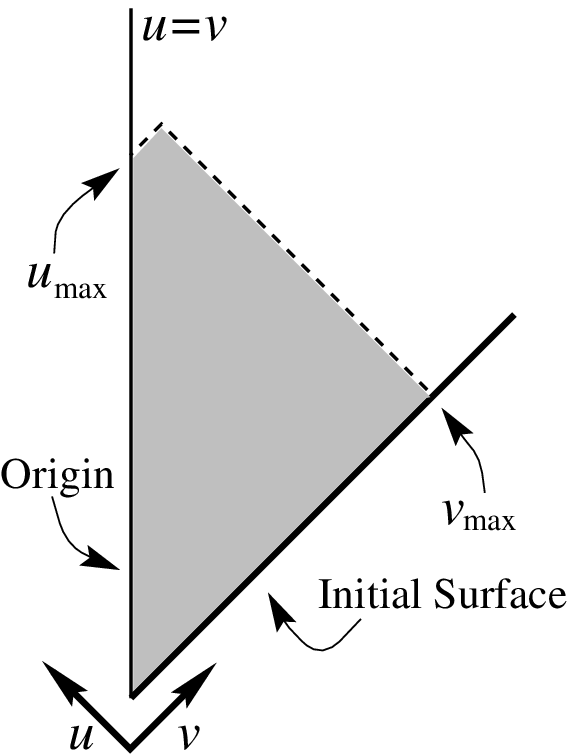}
  \includegraphics[width=3.9cm]{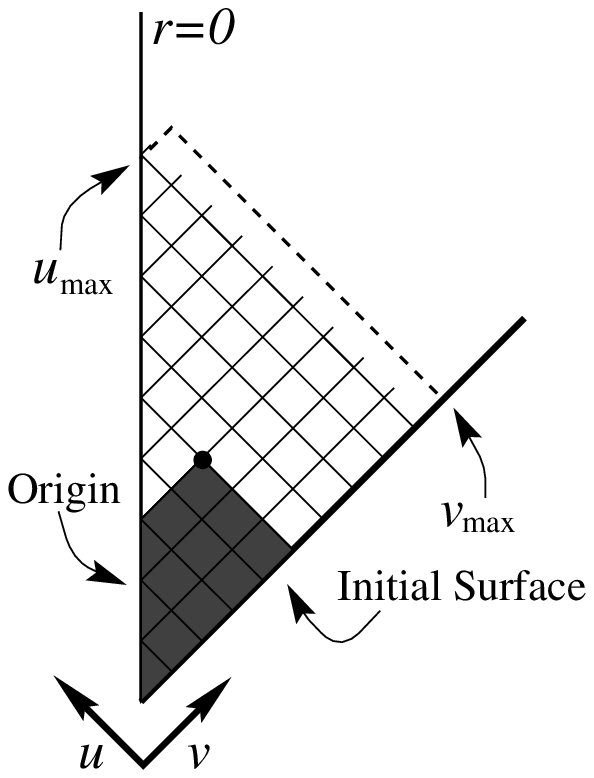} 
  b
  \caption[boundaries of integration]{\label{fig:boundary}
    (a) The boundaries of the computational region. Initial values are
    given on the initial surface and boundary conditions on the
    origin. The shaded area is the integration area. (b) The grid
    defined by double null coordinates. The casual past of some point
    $p$ is shaded }
\end{figure}
The characteristics of a massless scalar field are null so a double
null coordinate system $u,v$ is the "natural" system for this problem.
In these coordinates the horizon (when it forms) is regular and there
are no coordinate singularities. This enables us to study quantum
effects that occur near the horizon.

The  metric is:
\begin{equation}
        ds^2 = e^{2f}dudv-r^2d\Omega^2,
\end{equation}
Where $f$ and $r$ are functions of $u$ and $v$ only. This definition
is only unique up to a change of variables $v'=f_1(v), u'=f_2(u)$.
This ambiguity can be settled by a choice of the origin ($r=0$) as
$u=v$ and by choosing either $r$ or $f$ on the initial surface. We
choose $r=\frac{1}{2}v$ on $u=0$ which is supposed to be sufficiently
far in the past. This choice corresponds to an asymptotically flat
space-time. Regularity of the origin, with the above choice of
coordinates, implies the following boundary conditions:
\begin{mathletters}
\label{eq:bound_cond}
\begin{eqnarray}
        r_{,v}\Big|_{r=0} &=& -r_{,u}\Big|_{r=0},\\
        \phi_{,v}\Big|_{r=0} &=& \phi_{,u}\Big|_{r=0},\\
        f_{,v}\Big|_{r=0} &=& f_{,u}\Big|_{r=0}.
\end{eqnarray}
\end{mathletters}

The Classical Einstein equations together with the field's equation of
motion can be divided into two types: The dynamical equations
\begin{mathletters} 
\label{eq:kg}
\begin{eqnarray}
   r_{,uv} &=& 
        -\frac{\eta r}{2}, \\
   f_{,uv} &=& 
        -4\pi \phi_{,u}\phi_{,v} + \frac{\eta}{2}, \\
   \phi_{,uv} &=& 
        -\frac{1}{r}\left( {r_{,u}\phi_{,v} + r_{,v}\phi_{,u} }\right)
\end{eqnarray}
\end{mathletters}
and the constraint equations:
\begin{mathletters} 
\begin{eqnarray}
   -r_{,vv} + 2f_{,v}r_{,v} &=& 4\pi r\phi_{,v}^2 ,\label{eq:init}\\
   -r_{,uu} + 2f_{,u}r_{,u} &=& 4\pi r\phi_{,u}^2 .
\end{eqnarray}
\end{mathletters}

We have introduced here the auxiliary quantity $\eta$,
\begin{equation}
   \eta \equiv \frac{1}{r^2}\left(2r_{,u}r_{,v}+\frac{1}{2}e^{2f}\right) 
        = e^{2f}\frac{m}{r^3} \label{eq:hdef},
\end{equation}
where $m(u,v)$, is the mass inside a sphere of radius $r(u,v)$:
\begin{equation}
        m=\frac{r}{2}\left( 1+ 4r_{,u}r_{,v}e^{-2f}\right) 
        \label{eq:coeff_rel}.
\end{equation}
\section{Black Hole Evaporation \& The Semi-Classical Approximation}
We add to the classical equations an effective stress energy tensor
$\langle T^Q\rangle$ that describes the evaporation.  Following
\cite{piran} we turn to 2D theories to obtain a tensor embodying
Hawking radiation and adapt it to 4D. The expectation value of the 2D
quantum energy-momentum tensor \cite{davies}, in double-null
coordinates, is divided by $4\pi r^2$ to mimic the 4D radial
dependence to give:
\begin{equation}
{\langle T^Q\rangle}_{\mu\nu} = 
\frac{\alpha}{4\pi r^2}
\left [
 \begin{array}{cccc} 
  f_{,uu}-f_{,u}^2 & -f_{,uv} & 0 & 0 \\
  -f_{,uv} & f_{,vv}-f_{,v}^2 & 0 & 0 \\
  0 & 0 & 0 & 0 \\
  0 & 0 & 0 & 0
 \end{array}
\right ]
\end{equation}
The constant $\alpha$ defines a length scale $\sqrt{\alpha}$ which is
of the order of the Planck scale. Beyond this scale the semi-classical
approximation it is not valid.

Unfortunately this tensor diverges at the origin. Note that for a null
fluid considered by \cite{piran}, the origin is always flat and
$\langle T^Q\rangle$ vanishes identically there. Any attempt to
regularise this tensor must take into account that this tensor must be
divergence-less (i.e. $\nabla{\langle T^Q\rangle}=0$) because of the
Bianchi identity. A simple attempt to renormalize $\langle T^Q\rangle$
by multiplying it by a general function $Q(r)$ fails, we find that the
Bianchi identity is violated unless $Q={\rm Const}$.  Other attempts
which were based on an introduction of angular terms failed as well
and we had no choice but to ignore the Bianchi identity following
\cite{piran} and to multiply $\langle T^Q\rangle$ by
\begin{equation}
  Q = \frac{ \alpha/r^2} 
  {1+\left(\alpha/r^2\right)^2}. \label{eq:Q}\\ 
\end{equation}
The deviation from the Bianchi identity is significant only for $r
\lesssim \sqrt{\alpha}$, for which the whole the semi classical
approximation is no longer valid anyway.

The resulting semi-classical dynamical equations are:
\begin{mathletters}
\label{eq:dyn}
\begin{eqnarray}
   r_{,uv} &=& 
      \frac{r}{(1-Q)} \left( {4\pi Q\phi_{,u}\phi_{,v}-\frac{\eta}{2} }\right), \\
   f_{,uv} &=& 
      \frac{1}{(1-Q)} \left( { -4\pi \phi_{,u}\phi_{,v} + \frac{\eta}{2} }\right), \\
   \phi_{,uv} &=& 
      -\frac{1}{r}\left( {r_{,u}\phi_{,v} + r_{,v}\phi_{,u} }\right),
\end{eqnarray}
\end{mathletters}
and the constraint equations are:
\begin{mathletters}
\label{eq:constr}
\begin{eqnarray}
   -r_{,vv} + 2f_{,v}r_{,v} &=& 4\pi r\phi_{,v}^2 +
          Qr\left( { f_{,vv} - f_{,v}^2 }\right), \label{eq:bound}\\
   -r_{,uu} + 2f_{,u}r_{,u} &=& 4\pi r\phi_{,u}^2 +
          Qr\left( { f_{,uu} - f_{,u}^2 }\right). 
\end{eqnarray}
\end{mathletters}

\section{Numerical Scheme}
\begin{figure}
  \centering
  \noindent
  a 
  \includegraphics[width=4cm]{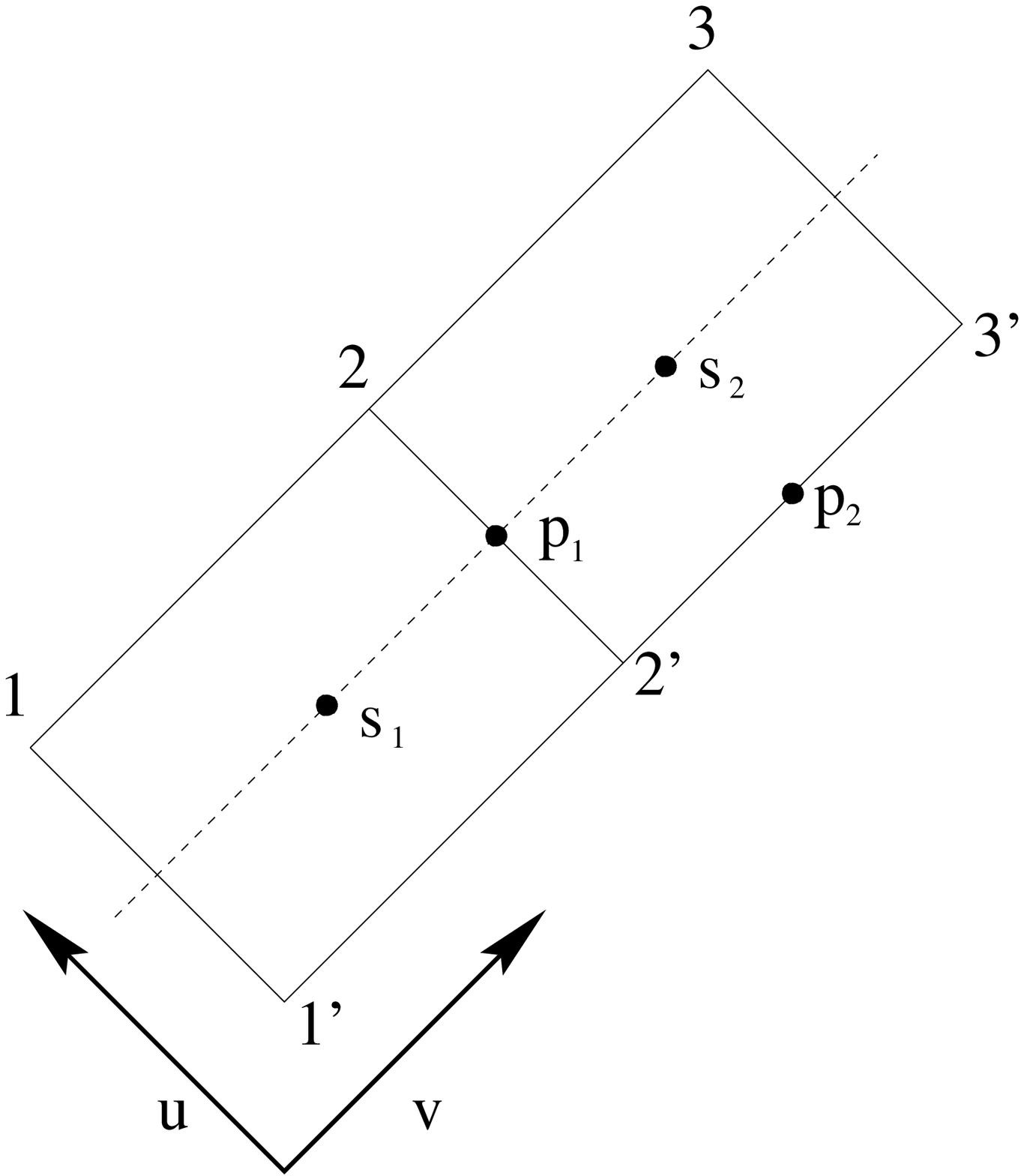} 
  \includegraphics[width=3.9cm]{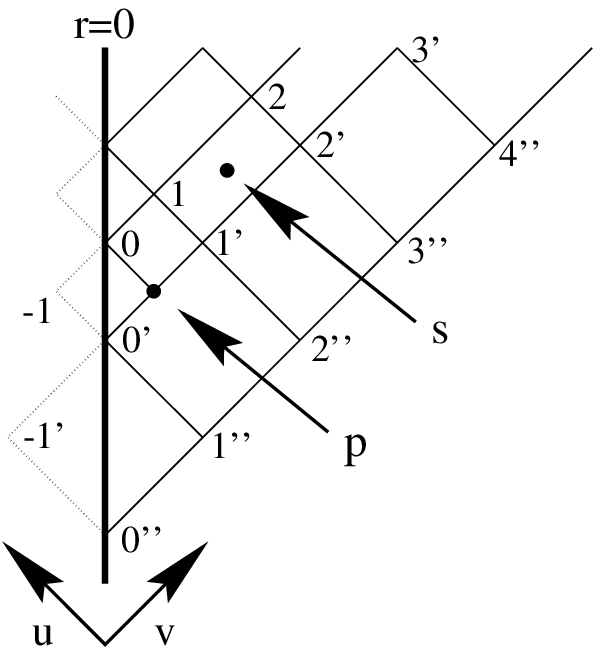}
  b
  \caption[Grid]{\label{fig:grid}
    (a) The regular grid structure away from the origin. The source
    terms are calculated at $\rm s_1$ and $\rm s_2$. $\eta$ is
    integrated along the dashed line and the derivatives are
    calculated at $\rm p_1$, $\rm p_2$ in the first iteration. (b) A
    typical grid arrangement near the origin. The point $p$ is
    interpolated.  Notice that $du\neq dv$ (e.g. cell 1-2-2'-1') and
    that the first rectangular cell might differ from the rest of the
    cells (e.g 0-1-1'-p as compared to 1-2-2'-1').  Points -1 and -1'
    are ``fake'' points calculated from the symmetry conditions.  }
\end{figure}
We choose $u=0$ as our initial surface and use the constraint equation
Eq.\ (\ref{eq:bound}) to determine the value of $f$ given $r$ and
$\phi$ on this line. The integration proceeds, for each constant $u$,
line from the origin $v=u$, where the boundary conditions Eq.\ 
(\ref{eq:bound_cond}) are imposed, up to $v_{\rm max}$ which is an
arbitrarily chosen maximal value of $v$. In this way, the complete
casual past of each point is known when the integration routine has to
determine the field's value at that point.

The straightforward numerical scheme becomes unstable once the
gravitational effects become important ($\frac{2m}{r}\sim 1$). To
stabilize the scheme, we have introduced the auxiliary quantity $\eta$
Eq.  (\ref{eq:hdef}) which is integrated on constant $u$ lines.  We
use this value of $\eta$ in the source for the evolution equations.
It is worth noting that $\eta$ needs to be accurate only to first
order since it is a source term. This modification stabilizes the
scheme. This is the first numerical scheme to solve this problem
using the evolution equations in double null coordinates.

A grid refinement algorithm was used in order to increase resolution at
specific places such as the near-critical solutions and at the
evaporation stage of the black-hole.

We take $u=0$ to be the initial surface. On this surface, we specify
an initial value for the scalar field $\phi$. Then we choose
$r=\frac{1}{2}v$ on the initial surface and solve Eq.\ 
(\ref{eq:bound}) for $f$.  The quantum terms should be negligible on
the initial surface, so they were ignored while solving for $f$. We
later verify that these terms are indeed negligible. The initial value
equation for $f$ is
\begin{equation}
   f_{,v} = 2\pi\frac{r}{r_{,v}}\phi_{,v}^2 
        = 2\pi v\phi_{,v}^2, \label{eq:bound_r}
\end{equation}
with $f(0,0)=0$ as a boundary condition. Eq.\ (\ref{eq:bound_r}) is
integrated using a fourth order Runge-Kutta algorithm.

Once the values of the fields on $u=0$ are known the integration
proceeds as follows: (all references to grid points and cells are
illustrated in Fig.\ (\ref{fig:grid}))

The first step involves a triangular cell on the origin. Here we
utilize the boundary conditions Eq.\ (\ref{eq:bound_cond}).  Because
space-time is flat near the origin, the $r$ direction is $(1,-1)$ when
taken as a vector in the $(v,u)$ plane. Thus we can approximate
$\frac{\partial \phi}{\partial r}$ to order $(\Delta r)^2$ in the
current step by using the value of $\phi$ at the two last steps.
\begin{itemize}
  
\item Using the value of $\phi_{1'}$ and $\phi_{3''}$ we obtain the
  value of $\phi_0$.
  
\item Once the value of $\phi_0$ is known, $\phi_{,v}$ and
  $\phi_{,uv}$ are calculated on the origin between the points 0
  and 0' using also $\phi_{-1}=\phi_{p}$ (again using
  Eq.\ (\ref{eq:bound_cond}) and $\phi_{0'}$).
  
\item Now we calculate the value of $f_{,uv} = \eta/2 = 4\pi/3
  \phi_{,v}^2$ and using the known $f_{0'}$ and $f_{-1}=f_p$ we obtain
  $f_0$
  
\item Finally we  calculate $f_{,v}$ at the same point as
  $\phi_{,v}$ and $\phi_{,uv}$ and from them $\eta$ and $\eta_{,v}$
  using Eq.\ (\ref{eq:h_orig}) and Eq.\ (\ref{eq:hvo}) (see appendix A
  for derivation of these equations).
\end{itemize}

All interpolations are done using a third degree polynomial such that
the point being interpolated always has two known points on each side
(e.g. the values $z_{-1'},z_{0'},z_{1'}$ and $z_{2'}$ are used to
interpolate $z_p$).

Away from the origin we have rectangular cells. For each step in the
$v$ direction three points are known (from previous steps) and the
fourth (the future-most) is calculated (e.g $z_2,z_{2'}$ and
$z_{3'}$ are known and $z_3$ is calculated). The integration is a
two step iterative procedure. The basic principle is that
the second derivative operator can be discretesized as follows:
\begin{equation}
        z_{,uv}\Big|_{s_2} = \frac{z_3-z_2-z_{3'}+z_{2'}}{\Delta u
                                        \Delta v} + O((\Delta u\Delta v)^2) 
            \label{eq:diffop}
\end{equation}
The source terms of the equations are split into two - the
gravitational term involving $\eta$ and the field term. The
gravitational term is integrated along a constant $u$ line which runs
through the point at which the source is calculated (in between of the
grid-lines). The field term is calculated directly.
\begin{description}
\item[$\bf 1^{\bf st}$ Iteration] The fields and their derivatives at
  point $s_2$ are approximated using the known values at 2,2' and 3':
\begin{eqnarray}
        z\Big|_{s_2} &\approx& \frac{1}{3}(z_2+z_{2'}+z_{3'}) \\
        z_{,v}\Big|_{s_2} &\approx& \frac{1}{\Delta v}(z_{3'}-z_{2'}) \\
        z_{,u}\Big|_{s_2} &\approx& \frac{1}{\Delta u}(z_{2}-z_{2'})
\end{eqnarray}
$\eta$ is integrated using $\eta_{,v}\Big|_{s_1}$ calculated for the
previous point. Eq.\ (\ref{eq:diffop}) yields a first approximation
for $z_3$.
\item[$\bf 2^{\bf nd}$ Iteration] A second order approximation for the
  fields and their derivatives can now be obtained using $z_3$ calculated
  in the first iteration:
\begin{eqnarray}
        z\Big|_{s_2} &\approx& \frac{1}{4}(z_2+z_3+z_{2'}+z_{3'}), \\
        z_{,v}\Big|_{s_2} &\approx& 
                \frac{1}{2\Delta v}(z_{3'}-z_{2'} + z_{3}-z_{2}), \\
        z_{,u}\Big|_{s_2} &\approx& 
                \frac{1}{2\Delta u}(z_{2}-z_{2'}+z_{3}-z_{3'}).
\end{eqnarray}
An improved approximation for $\eta$ is given by:
\begin{equation}
        \eta_{,v}\Big|_{p_1} \approx \frac{1}{2}
                \left(\eta_{,v}\Big|_{s_1}+\eta_{,v}\Big|_{s_2} \right).
\end{equation}
\end{description}
This iterative explicit scheme gives us a second order accuracy.

To study the last stages of collapse, we utilize a cell refinement
algorithm that increases the resolution of the integration in the
areas of interest. Various \emph{multi grid} or \emph{adaptive mesh}
\cite{berger} methods have been used to study critical collapse (cf.
\cite{chop} and \cite{stewart}) but the increased complexity of these
schemes was unnecessary for our purpose. We chose to half the value of
$\Delta u$ at preset intervals, keeping the value of $\Delta v$ constant.
This scheme has the advantage that we can reach an asymptotic value of
$u$ which is independent of the value of $v_{\rm max}$. Using it we
can study structure away from the origin. The price is that the first
rectangular cell can be different from the others (see Fig.{}
\ref{fig:grid}b). This scheme also sacrifices $v$ resolution. Both
problems are not critical.
\section{Results}
\begin{figure}
  \begin{center}
    \leavevmode
    a
    \includegraphics[width=6cm]{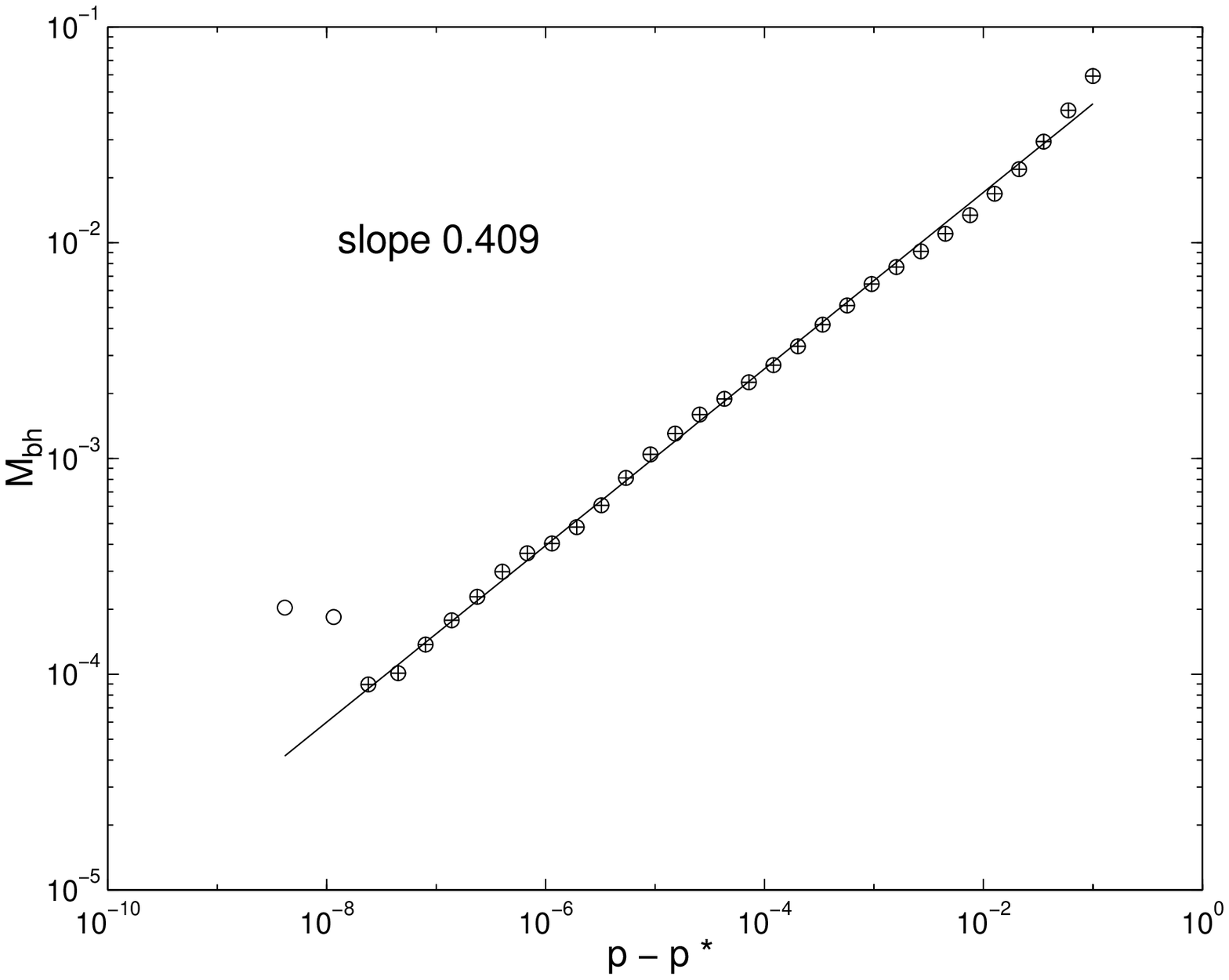}
    \includegraphics[width=6cm]{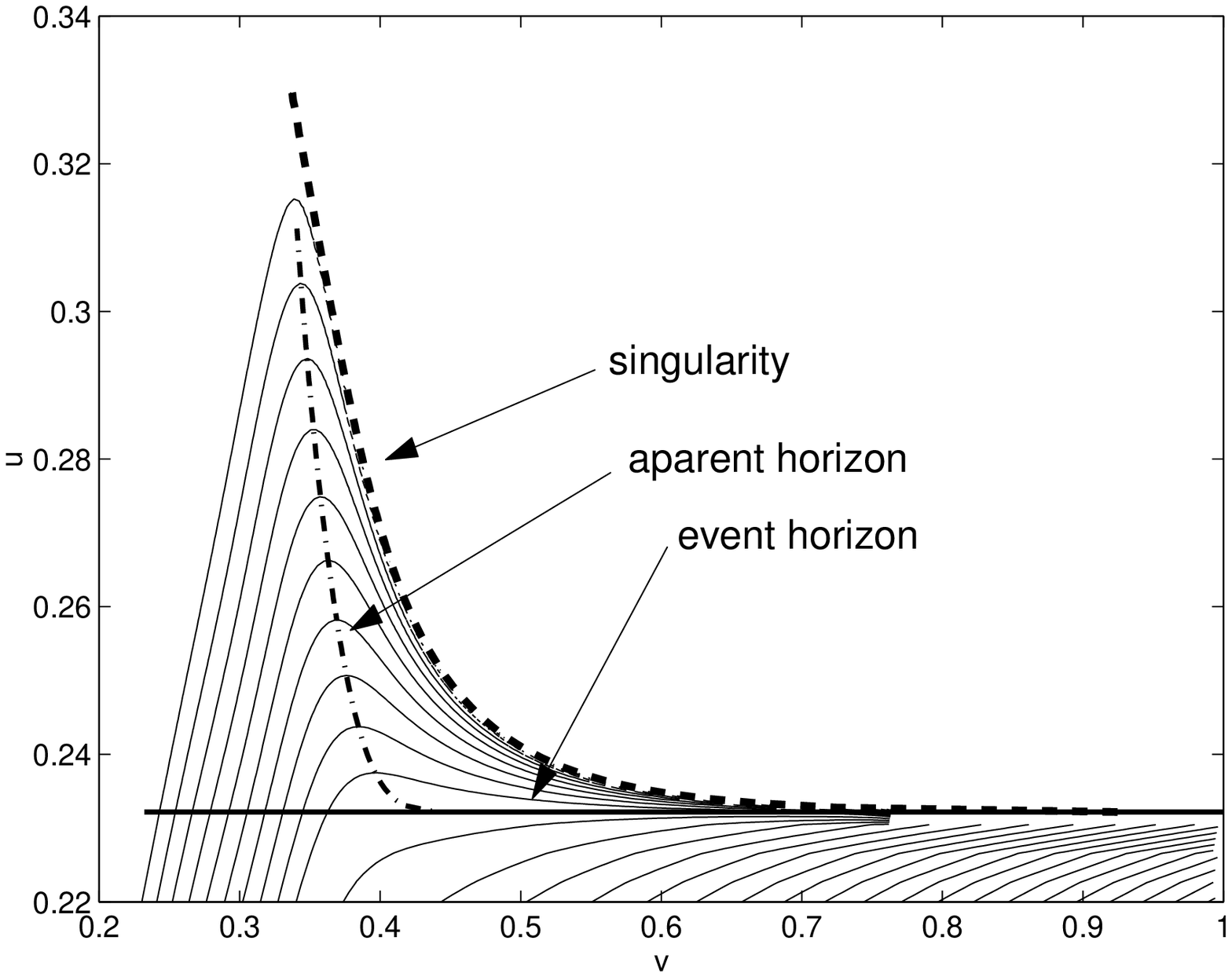}
    b
    \caption[Classical results]{
      Classical results: (a) A log-log plot of $M_{\rm bh}$ vs.
      $p-p^*$. The slope, 0.409, is higher then the usually quoted
      result of 0.375. Both the slope and $p^*$ were found through
      least squares fitting. (b) A contour plot of $r$ values in the
      case of formation of a black hole. The event horizon, apparent
      horizon \& the singularity are marked. The apparent horizon
      is always beyond (larger $u$) the event horizon. }
    \label{fig:class}
  \end{center}
\end{figure}
To test our code we first ran it on the classical ($\sqrt{\alpha}=0$)
problem.  We find a critical initial amplitude $p^*$. For initial data
with amplitudes smaller then $p^*$ the field disperses leaving a flat
space-time. For initial amplitudes above $p^*$ the field collapses to
form a black hole. For amplitudes near $p^*$ the solution exhibits
self similar oscillations. A log-log plot of $M_{\rm bh}$ as a
function of $p-p^*$ can be seen in Fig.\ (\ref{fig:class}). Although
the critical exponent we found, $\gamma=0.409$ is larger then the
usually quoted number $\gamma \approx 0.375$ the power law dependence
of $M_{\rm bh}$ on $p-p^*$ is evident. Thus our code reproduces the
classical behavior. The discrepancy in the value of $\gamma$ reflects
the accuracy of the code. We also present a contour plot of $r$
showing the apparent horizon, the singularity and the event horizon
(defined as the last ray to avoid the singularity). For the classical
case, the apparent horizon is always inside the event horizon - any
photon which starts falling into the black hole will never escape.

The introduction of $\langle T^Q\rangle$ creates an effective outgoing
flux corresponding to Hawking radiation.  This flux produces an
evaporation satisfying Eq.\ (\ref{eq:evapor}) as verified in
\cite{piran} for black holes created by a null fluid.  It is sometimes
hard to distinguish the evaporation from the energy momentum of the
scalar field which is reflected from the origin. These reflections and
the tail of the incoming scalar field can mask the quantum flux. To
see a distinct black hole evaporation we need to create a situation
where the only energy flux crossing the horizon is the outgoing
Hawking radiation. This can be accomplished by looking at initial
conditions that create a black hole before a significant fraction of
the scalar field's energy is reflected from the origin.  We also look
for conditions producing an apparent horizon after most of the field
has collapsed (large $v$) so that the tail of the field is also inside
the horizon.

\begin{figure}
  \centering
  \noindent
  a\\
  \includegraphics[clip,width=10cm]{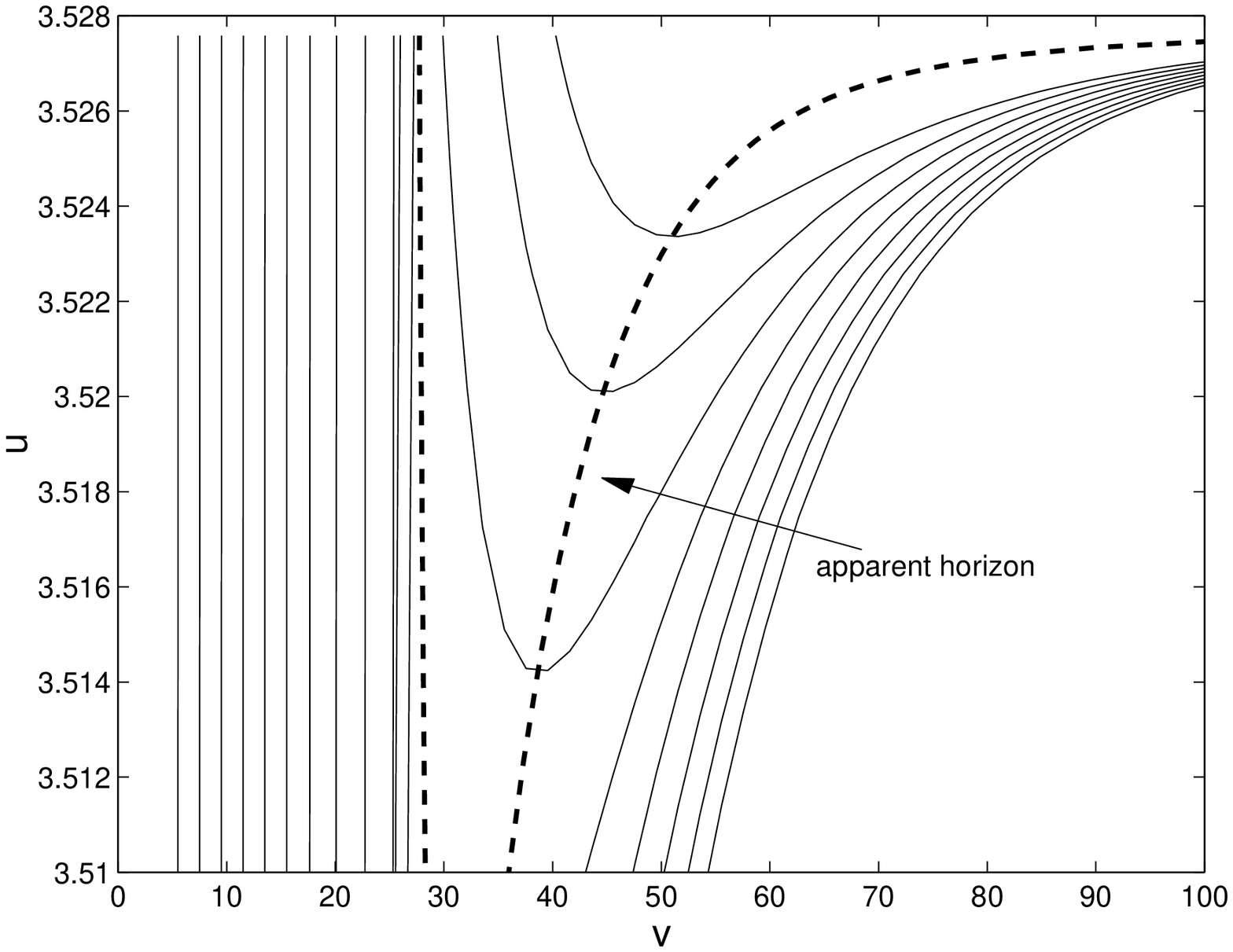} \\
  b\includegraphics[clip,width=6cm]{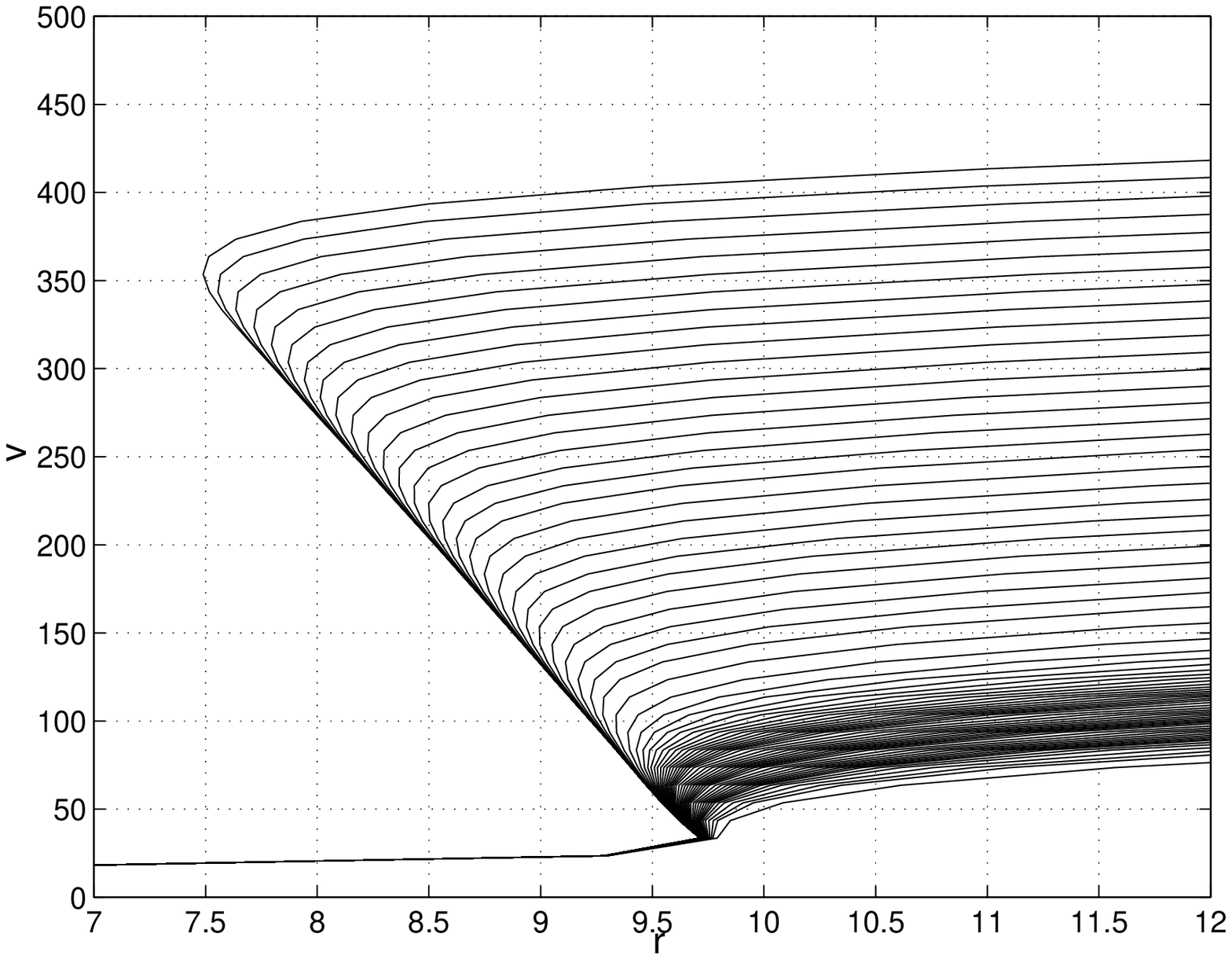}
  c\includegraphics[clip,width=6cm]{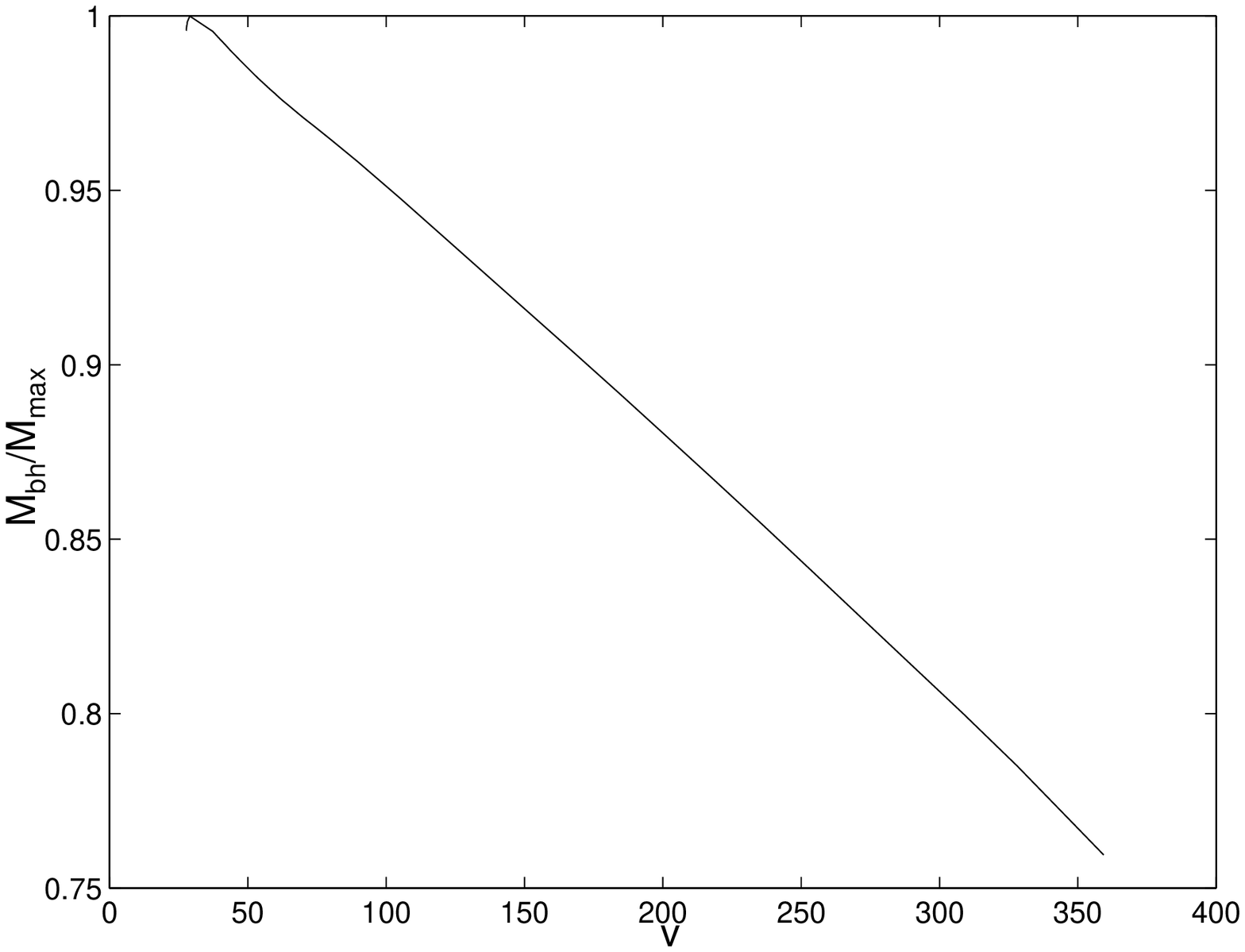}
  \caption[Black hole evaporation]{\label{fig:ev_cont}
    (a) Contour of $r$ in the evaporation region. The field is
    centered around $v=20$, to the left of it the space time is flat
    ($r \propto v-u$). The dotted line describes $r_{,v}=0$ - the
    apparent horizon. The event horizon is at the top of the graph
    ($u=3.528$). (b) trajectories of $r(v)$ for different values of
    $u$. These null trajectories curve back towards the singularity
    but then turn around and escape to infinity. This is a clear
    indication for mass decrease. The decrease in radius of the
    apparent horizon is evident.  (c) The black hole's mass as
    function of $v$. A 25\% decrease in mass is evident.}
\end{figure}
Fig.\ (\ref{fig:ev_cont}) depicts the geometry resulting from the
evolution of a scalar field with such initial conditions. In the $r$
contour plot Fig.\ (\ref{fig:ev_cont}a) The appearance of the apparent
horizon before (lower $u$) the event horizon is apparent. Null
trajectories which begin to curve back towards the singularity but
then turn around and escape to infinity are shown in the $r(v)$ plot
Fig.\ (\ref{fig:ev_cont}b). The evaporation of the black hole is
indicated by the decreasing radius of the outer apparent horizon which
implies a negative energy flux through the apparent horizon.

It is difficult to resolve the evaporation in the $u,v$ coordinates
which constitute the inertial frame of an observer at rest on the
origin.  In these coordinates, the evaporation is contained within a
tiny $u$ lapse. We therefore utilize the redshift of outgoing rays (as
in \cite{piran}) which implies that $r_{,u}\Big|_{r \gg m }$ diverges
as the event horizon is approached. We half the $u$ step $\Delta u$ in
order to always have:
\begin{equation}
  r_{,u}\Delta u \Big|_{v=v_{\rm max}} < \Delta r, \label{eq:dr}
\end{equation}
where $\Delta r$ is the value of the lhs. of Eq. (\ref{eq:dr}) for the
first $u$ step. Using this scheme we managed to resolve a 25\%
decrease in the black hole mass due to evaporation.

In our search for the scaling law which characterizes the Choptuik
phenomena, we have to stop when the resulting black hole mass
approaches $\sqrt{\alpha}$. Nevertheless, we can still take $\alpha
\ll 1$ and try to get a scaling law down to the minimal meaningful
mass.

\begin{figure}
  \centering 
  \noindent
  \includegraphics[clip,width=6cm]{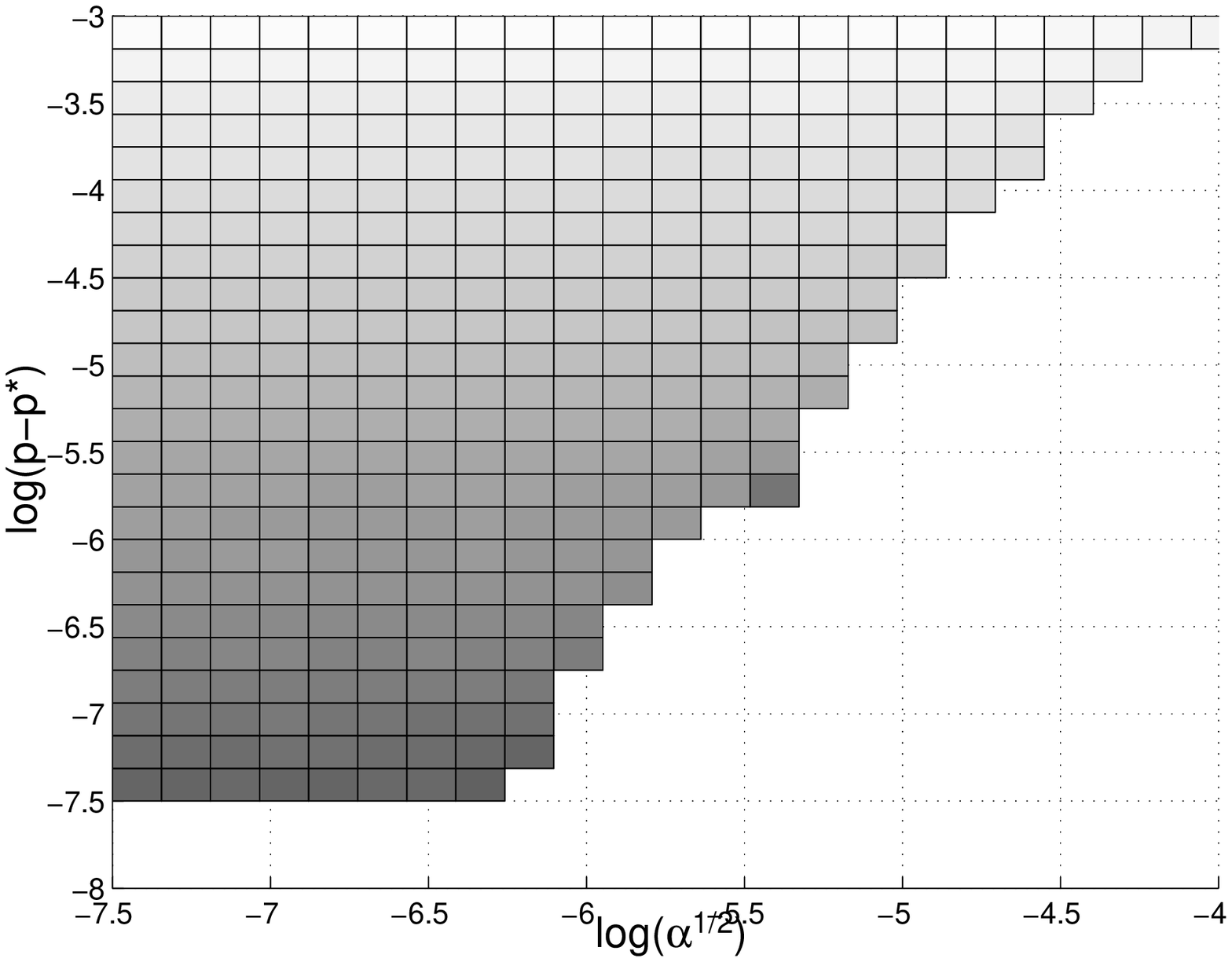}
  \includegraphics[clip,width=6cm]{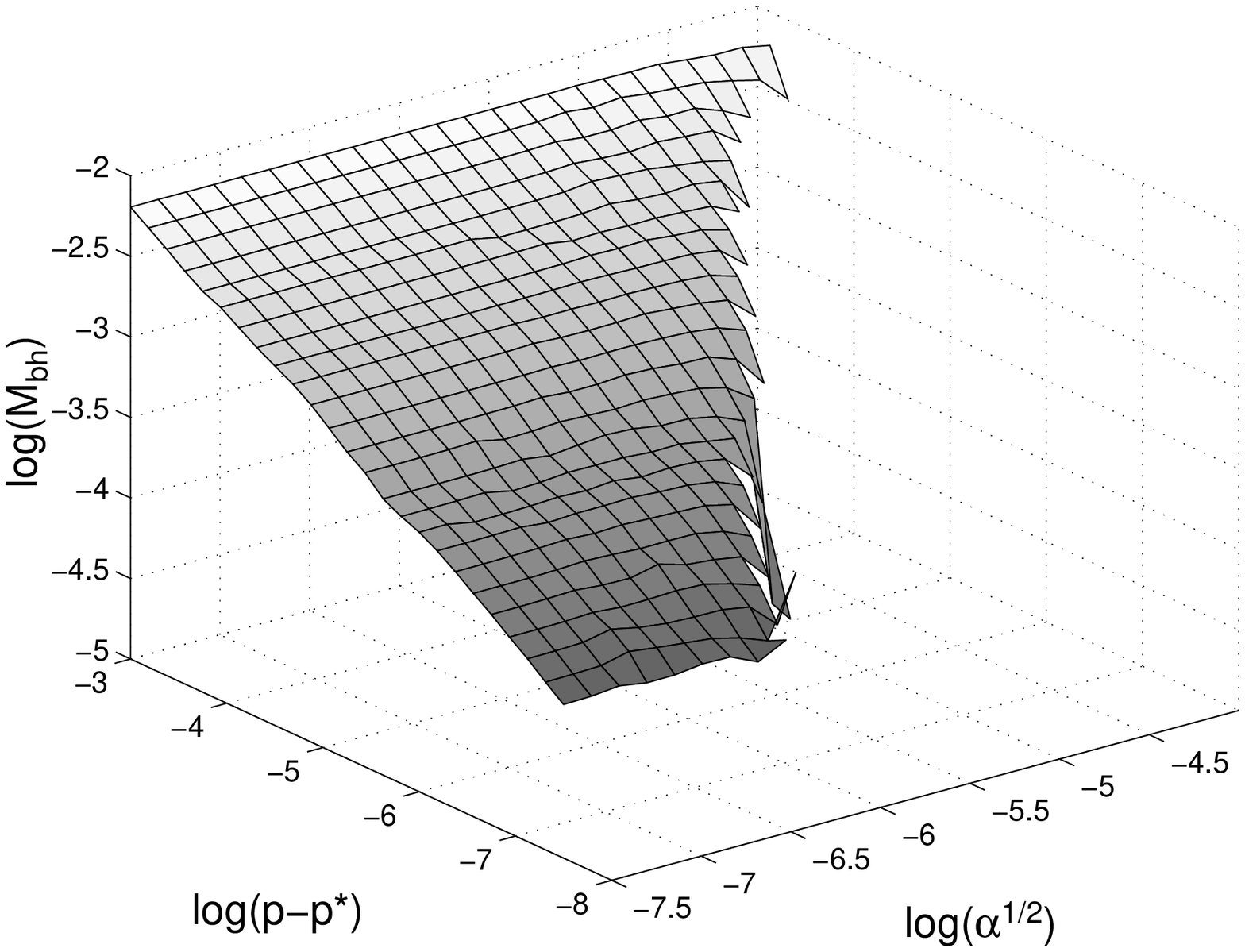}
  \caption[M as function of $\alpha$ and $p-p^*$]{\label{fig:p-a_surf}
    $M_{\rm bh}$ as function of $\sqrt{\alpha}$ and $p-p^*$ Darker
    colors indicate lower mass. As $\sqrt{\alpha}$ increases $M_{\rm
      bh}$ decreases until finally no black hole is formed. The empty
    region is where no black holes formed. The final decrease in $M_{\rm
      bh}$ is too rapid to be resolved in this graph }
\end{figure}
\begin{figure}
  \centering
  \noindent
  \includegraphics[width=4cm,angle=-90]{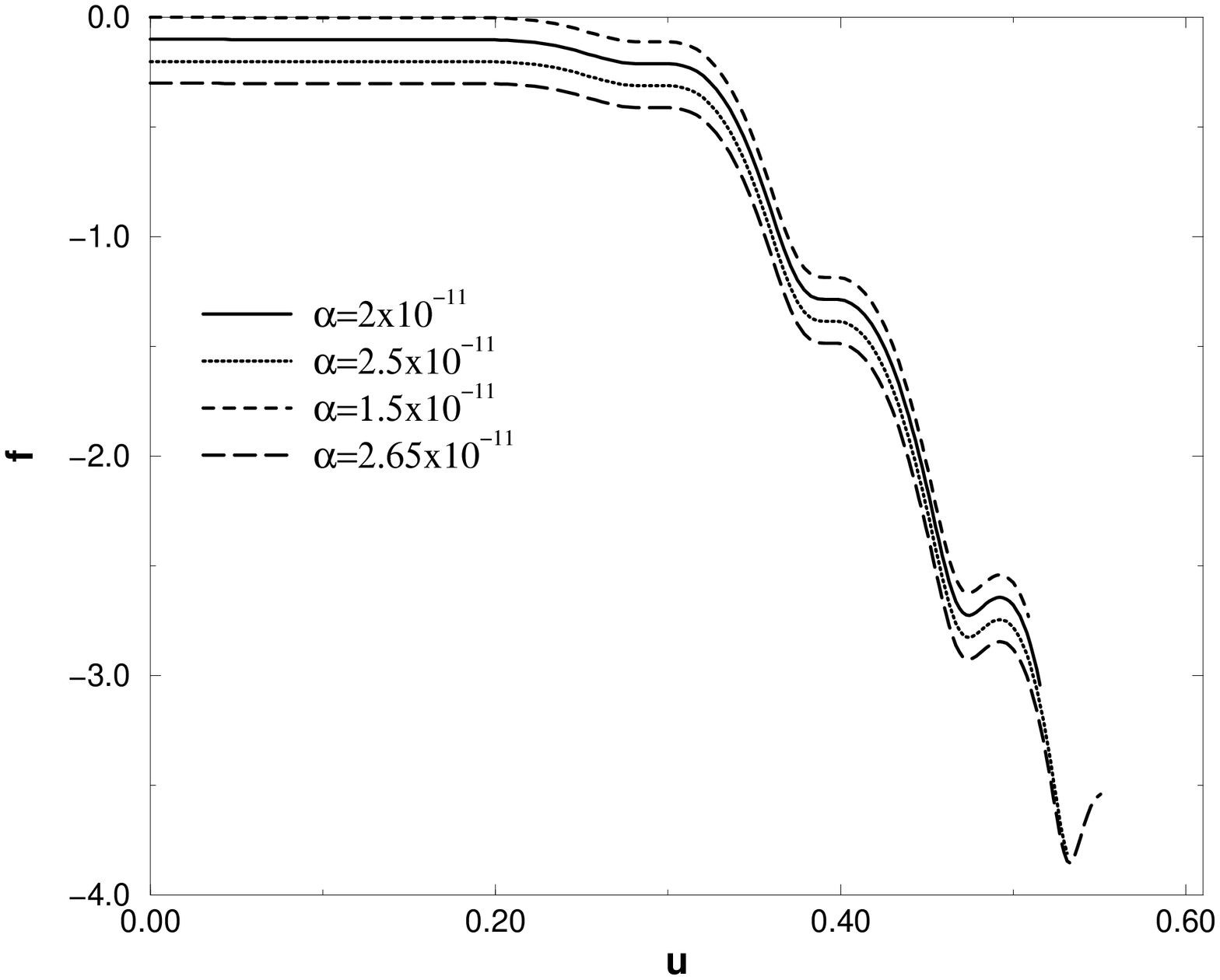}
  \includegraphics[width=4cm,angle=-90]{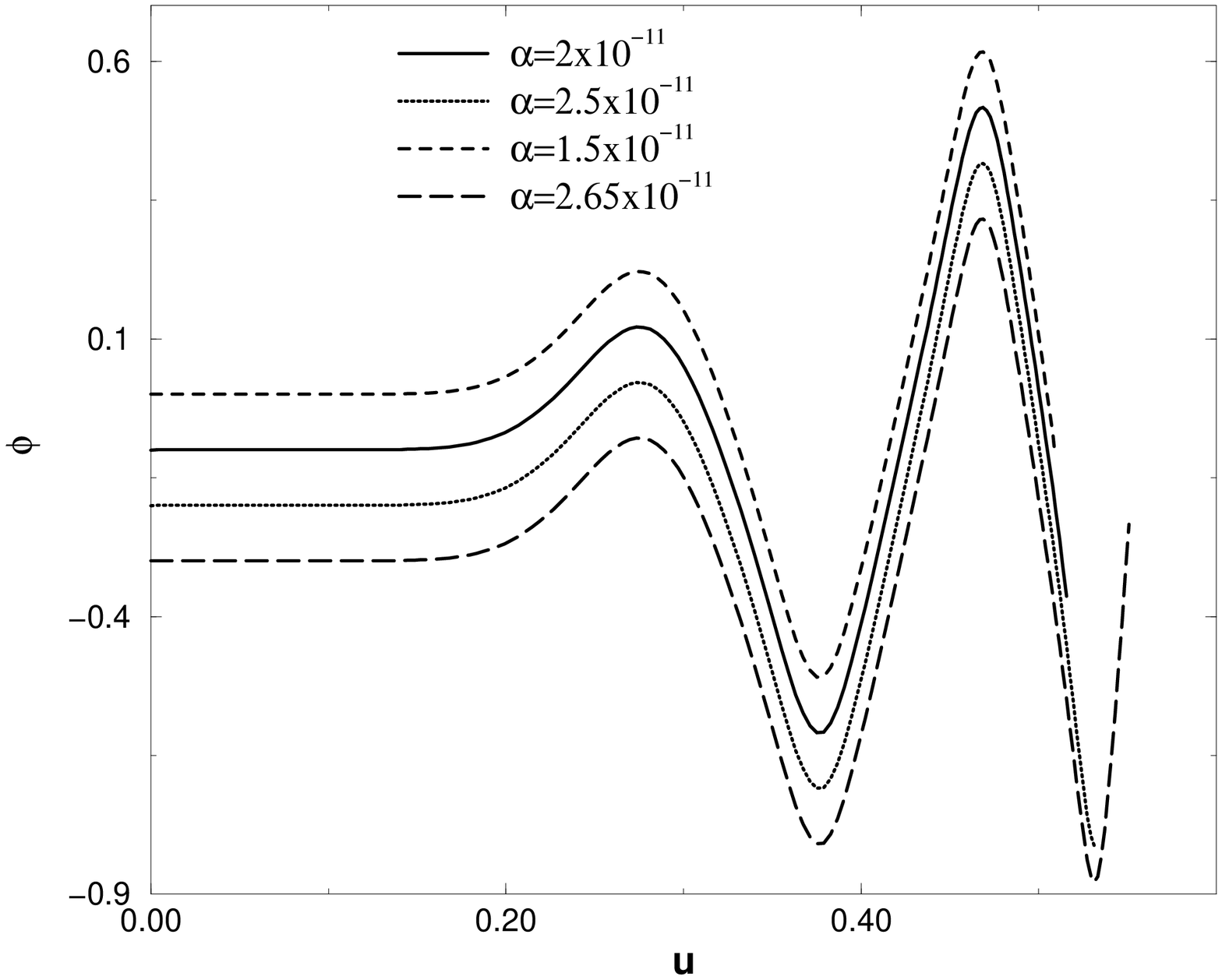}
  \caption[Fields on origin]{\label{fig:origin_fld}
    f and $\phi$ on origin for fixed initial data (fixed $p$) and
    different $\sqrt{\alpha}$. The different curves are offset vertically in order
    to enhance readability. Increasing $\sqrt{\alpha}$ increases the number of
    echoes and is thus equivalent to decreasing $p$ }
\end{figure}

In Fig.\ (\ref{fig:p-a_surf}) we show a plot of $M_{\rm bh}$ as a
function of $p-p^*$ and $\sqrt{\alpha}$. For small $\sqrt{\alpha}$ the
evaporation is negligible and we recover the power law dependence of
$M_{\rm bh}$ on $p-p^*$. As we increase $\sqrt{\alpha}$ the effects of
mass loss is apparent. For large values of $\sqrt{\alpha}$ black holes
do not form where they would have classically.  Thus, increasing
$\alpha$ is effectively equivalent to decreasing $p$.  This occurs
because of mass loss during the evolution. These losses by evaporation
can even change supercritical classical initial data into a
subcritical solution and inhibit the formation of a black hole.
\begin{figure}
  \centering
  \noindent
  a\includegraphics[width=6cm]{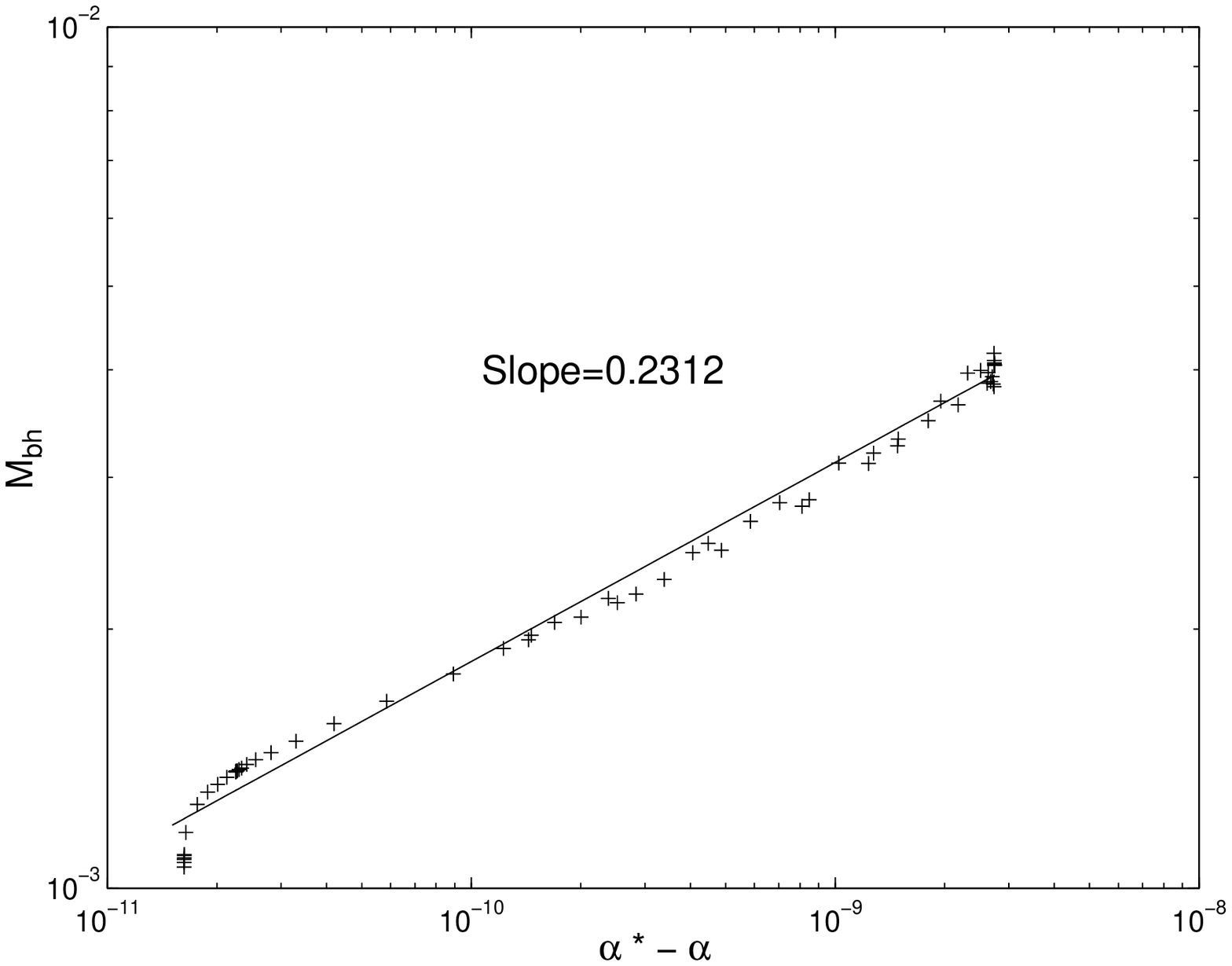}\\
  b\includegraphics[width=6cm]{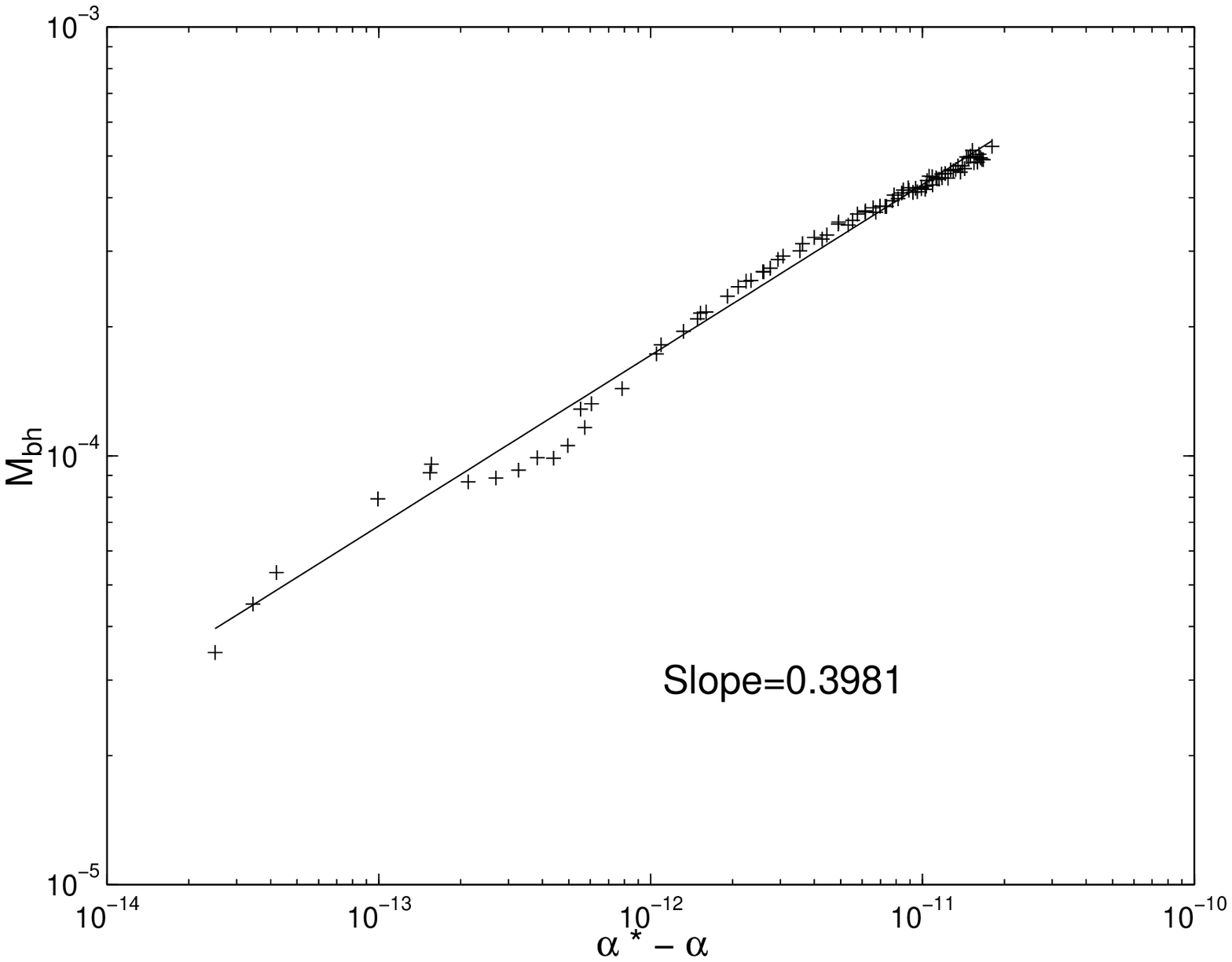}\\
  c\includegraphics[width=6cm]{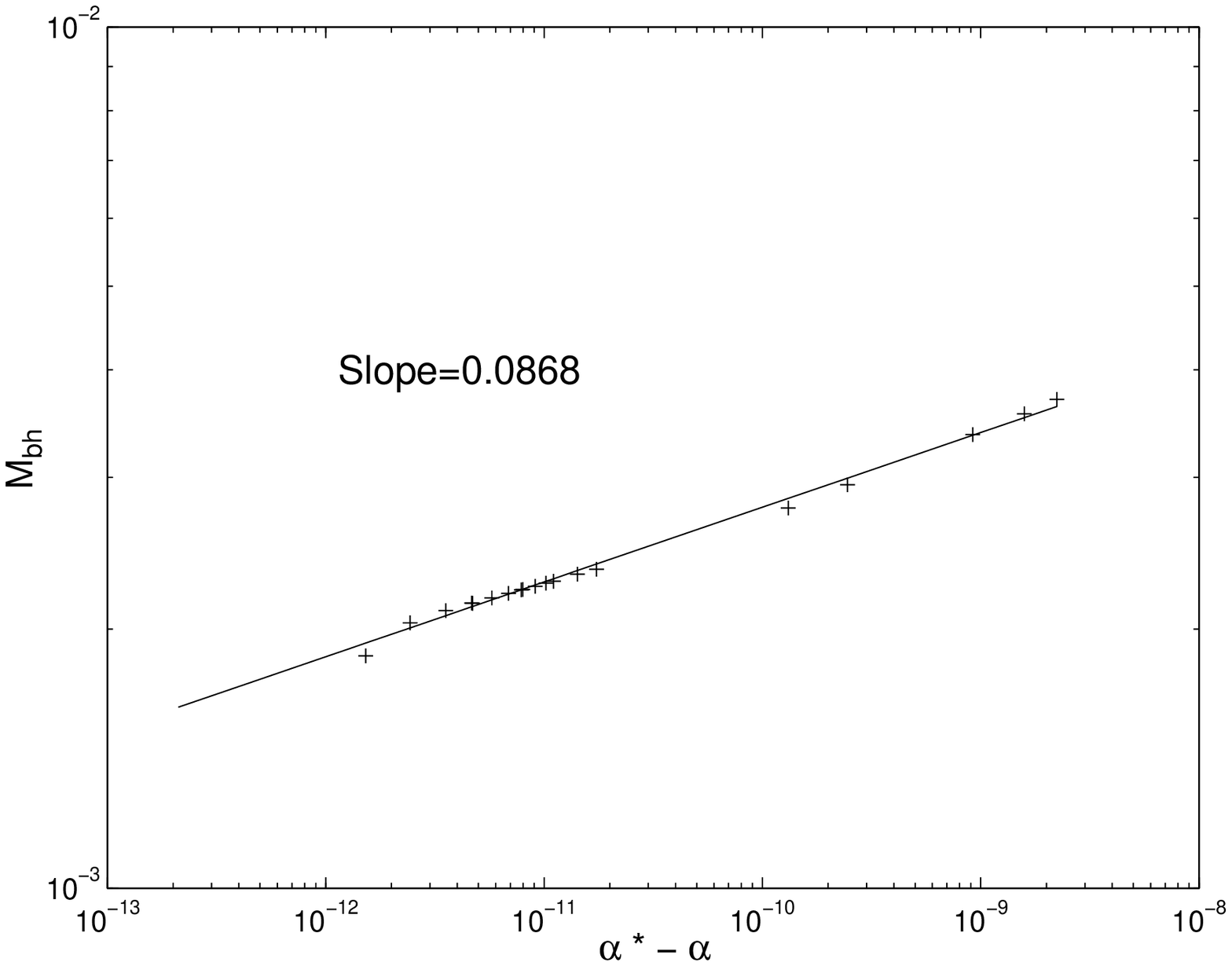}
  \caption[$M_{\rm bh}$  as function of $\alpha^*-\alpha$ for different $p$ ]{
    \label{fig:p-a_selected} $M_{\rm bh}$  as function of $\alpha^*-\alpha$ for
    three different $p$. $\alpha^*$ was determined through fitting.
    The values of $\alpha^*$ are $3.8\times 10^{-8}$, $2.8\times
    10^{-9}$ and $2.7\times 10^{-11}$ for graphs (a), (b) and (c)
    respectively.  }
\end{figure}
The decrease in mass with increasing $\sqrt{\alpha}$ is too rapid to
be resolved by the coarse grid of Fig.\ (\ref{fig:p-a_surf}). In
Figs.\ (\ref{fig:p-a_selected}) and (\ref{fig:origin_fld}) we examine
more closely the changes in the evolution due to increasing
$\sqrt{\alpha}$. In Fig.\ (\ref{fig:p-a_selected}) we show, $M_{\rm
  bh}$ as a function of $\sqrt{\alpha}$ for three different values of
$p-p^*$. A power law dependence of $M_{\rm bh}$ on $\sqrt{\alpha}$
with an exponent which is dependent on the value of $p-p^*$ can be
seen.  The power law depicted in Fig.\ (\ref{fig:p-a_selected}) is
broken before the semi-classical limit is reached by a rapid fall in
$M_{\rm bh}$.  This feature is also present in the classical
($\alpha=0$) case indicating that we have reached our limiting
resolution.
Fig.\ (\ref{fig:origin_fld}) depicts $\phi$ and $f$ on the origin for
various values of $\sqrt{\alpha}$ with a fixed value of $p$. We see
that the effect of increasing $\sqrt{\alpha}$ is practically similar
to that of decreasing $p$. The general echoing structure of near
critical solutions is conserved and the decrease in mass affects only
the last echo. This is in agreement with the power law dependence of
$M_{\rm bh}$ on $\sqrt{\alpha}$ since each echo is exponentially
smaller then the preceding one in mass.

\section{Summary}
The numerical solution of spherical symmetric semi-classical scalar
field collapse enabled us to study the effects of black hole
evaporation on the critical phenomena present in classical collapse
(the Choptuik effect). The code was tested to reproduce the critical
phenomena in the classical limit and black hole evaporation in the
semi classical approximation.

The addition of quantum effects and the subsequent introduction of the
Planck scale $\sqrt{\alpha}$ changed the behavior of the solutions.
For initial data found to form a black hole classically ($\alpha=0$)
the introduction of $\sqrt{\alpha}$ caused a decrease in the resulting black
hole mass $M_{\rm bh}$ . For large enough $\sqrt{\alpha}$, no black
hole formed. This behavior is expected due to mass loss through
evaporation. A detailed investigation of the relations between the
parameters revealed that for a fixed initial data (fixed $p$) there is a
power law dependence of $M_{\rm bh}$ on $\sqrt{\alpha}$. The exponent
of the power law depends on $p$. Unlike some earlier expectations
\cite{chiba} we find that the introduction of the Planck scale does
not destroy altogether the DSS behavior observed classically.
 
\appendix
\section{Auxiliary Quantities}
In order to carry out the numerical integration of the equations
(\ref{eq:dyn}) we need to calculate some auxiliary quantities, mainly
get an equation for $\eta_{,v}$ and get the value of $\eta$ and
$\eta_{,v}$ on the origin.

Using Eq.\ (\ref{eq:coeff_rel}) we can get an equation for $m$ from the
constraint equation Eq.\ (\ref{eq:bound})
\begin{eqnarray}
   2r_{,v}m_{,v} &=& 4\pi\left({ 1-\frac{2m}{r} }\right)r^2\phi_{,v}^2
\label{eq:quant} \\
      &+& \left({ 1-\frac{2m}{r} }\right)r^2 Q \nonumber\\
          &\times& \left[{
         \left({f_{,vv}-f_{,v}^2 }\right) - \frac{1}{1-Q}\left({
            4\pi\phi_{,u}\phi_{,v} + \frac{2m}{r^3}\frac{r_{,v}}{ 1-\frac{2m}{r}}
         }\right)
      }\right] \nonumber 
\end{eqnarray}
using this we can calculate $\eta$ on the origin (which is always
classic)
\begin{equation}
     m(u,v) = \int_u^{u+v} 2\pi\left(
        { 1-\frac{2m}{r} }\right)\frac{r^2}{r_{,v}}\phi_{,v}^2 dv'
\end{equation}
so by a change of variable from $v$ to $r$ at constant $u$ we have
\begin{equation}
     m\left({u,r(u,v)}\right) = \int_0^r 2\pi\left({ 1-\frac{2m}{r'} }\right)
     \frac{{r'}^2}{{r'}_{,v}^2}\phi_{,v}^2 dr' 
\end{equation}
now, to the lowest order in $r$, we have (since $\frac{2m}{r}\propto
o\left({r^2}\right)$)
\begin{equation}
     m = 2\pi\frac{\phi_{,v}^2}{r_{,v}^2} \int_0^r {r'}^2 dr'
\end{equation}
where all the terms out of the integral are to be evaluated at $r=0$.
Using $r_{,v}^2=e^{2f}/4$ on the origin, we have, for small $r$
\begin{equation}
     m \approx \frac{8\pi}{3}\phi_{,v}^2 e^{-2f} r^3
\end{equation}
and so, on the origin, using Eq.\ (\ref{eq:hdef})
\begin{equation}
     \eta = \frac{8\pi}{3}\phi_{,v}^2 \label{eq:h_orig}
\end{equation}
to get an equation for $\eta_{,v}$, using, again, Eq.\ (\ref{eq:hdef})
we have
\begin{equation}
     \eta_{,v}=\frac{e^{2f}}{r^3}\left({ 2mf_{,v} - 3m\frac{r_{,v}}{r} + m_{,v}}\right)
\end{equation}
with Eq.\ (\ref{eq:hdef}) and the quantum corrections from Eq.\ 
(\ref{eq:quant}) the equation is
\begin{eqnarray}
  \eta_{,v}     
  &=&
  \left({ 2f_{,v}-3\frac{r_{,v}}{r} }\right) \eta + 
  \frac{2\pi}{r_{,v}}\left({\frac{e^{2f}}{r}- 2\eta r}\right)\phi_{,v}^2
  \label{eq:hv} \\
  &+& 
  Q\left({\frac{e^{2f}}{r}- 2\eta r}\right) \nonumber\\
  &\times&
  \left[{
      \left({f_{,vv}-f_{,v}^2 }\right) - \frac{1}{1-Q} \left({
          4\pi\phi_{,u}\phi_{,v} + \frac{\eta r_{,v}}{e^{2f}-2\eta r^2}
          }\right)
      }\right]\nonumber 
\end{eqnarray}
We can find the value of the rhs. of Eq.\ (\ref{eq:hv}) on the origin
by expanding all the fields in powers of $v-u$. We then obtain the
expansion coefficients (which depend on $u$) by substituting back into
the Einstein equations Eq.\ (\ref{eq:dyn}). This is a rather tedious
process but can be easily automated using symbolic math programs such
as Mathematica \& Maple. By comparison of coefficients we deduce that:
\begin{equation}
  \eta_{,v}\Big|_{u=v}
  = \frac{8\pi}{3}\phi_{,v}\left({
      3\phi_{,uv}-2\phi_{,v}f_{,v} }\right) \label{eq:hvo}
\end{equation}
(all quantities are evaluated at the origin).
 
%
%




\begin{thebibliography}{}

\bibitem{christ} Christodoulou D., 
  Commun. Math. Phys.  {\bf 109}, 613 (1987) 
\bibitem{goldwirth} Goldwirth D. S., Piran T., 
  Phys. Rev. {\bf D36}, 3575 (1987) 
\bibitem{chop} Choptuik M. W., 
  Phys. Rev. Lett. {\bf 70}, 9 (1993) 
\bibitem{nrg} Hirschmann E., Eardly D., 
  Phys. Rev. {\bf D52}, 5850 (1995)
\bibitem{stewart} Hamad\'{e} R. S., Stewart J. M.,
  Class. Quantum Gravity {\bf 31}, 1 (1996) 
\bibitem{hawking} Hawking S. W., 
  Nature {\bf 248}, 30 (1974) 
\bibitem{page} Page D. N., 
  Phys. Rev. {\bf D13}, 198 (1976) 
\bibitem{semiclass} Birrell N., Davies P. W. C., 
  {\em Quantum Fields in Curved Spaces} 
  (Cambridge University Press, Cambridge, 1972) 
\bibitem{piran} Parentani R., Piran T., 
  Phys. Rev. Lett. {\bf 73}, 2805 (1994)
\bibitem{garf} Garfinkle D., 
  preprint gr-qc/9412008
\bibitem{chiba} Chiba T., Siino M.
  preprint {\bf KUNS 1384} (1996)
\bibitem{davies} Davies P. C. W., Fulling S. A., Unruh W. G., 
  Phys. Rev {\bf D13}, 2720 (1976)
\bibitem{berger} Berger M. J., Oliger J., J. 
  Comput. Phys. {\bf 53},  484 (1984) 
\end{thebibliography}
\end{document}